\PassOptionsToPackage{hyphens}{url}
\PassOptionsToPackage{usenames,dvipsnames,svgnames,table}{xcolor}
\documentclass[sigconf, letterpaper, screen=true, authorversion]{acmart}

\usepackage{url}
\usepackage{booktabs}
\usepackage{tabularx,colortbl}
\usepackage{makecell}
\usepackage{tikz}
\usepackage{layouts}
\usepackage{amsmath}
\usepackage{acronym}
\usepackage[unicode]{hyperref}
\usepackage{printlen}
\usepackage{units}
\usepackage{multirow}
\usepackage{multicol}
\usepackage{makecell}
\usepackage{textcomp}
\usepackage{paralist}

\usepackage{caption}
\usepackage{subcaption}

\newcommand{\eg}{e.g., }

\newcommand{\ie}{i.e., }

\newcommand{\etal}{et~al.}

\newcommand{\afblock}[1]{\noindent{\textbf{#1}}}
\newcommand{\h}[1]{H#1}

\usepackage{enumitem}
\setitemize{noitemsep,topsep=3pt,parsep=3pt,partopsep=3pt}

\usepackage{balance}

\acrodef{alpn}[ALPN]{Application Layer Protocol Negotiation}
\acrodef{npn}[NPN]{Next Protocol Negotiation}

\acrodef{plt}[PLT]{Page Load Time}

\acrodef{hol}[HoL]{Head of Line}

\acrodef{dom}[DOM]{Document Object Model}
\acrodef{cssom}[CSSOM]{CSS Object Model}

\acmYear{2018}\copyrightyear{2018}
\setcopyright{acmcopyright}
\acmConference[CoNEXT '18]{CoNEXT '18: International Conference on emerging Networking EXperiments and Technologies}{December 4--7, 2018}{Heraklion, Greece}
\acmBooktitle{CoNEXT '18: International Conference on emerging Networking EXperiments and Technologies, December 4--7, 2018, Heraklion, Greece}
\acmPrice{15.00}
\acmDOI{10.1145/3281411.3281434}
\acmISBN{978-1-4503-6080-7/18/12}

\date{}

\title{Is the Web ready for HTTP/2 Server Push?}

\author{Torsten Zimmermann, Benedikt Wolters, Oliver Hohlfeld, Klaus Wehrle}
\affiliation{
  \institution{Communication and Distributed Systems, RWTH Aachen University}
}
\email{{zimmermann, wolters, hohlfeld, wehrle}@comsys.rwth-aachen.de}

\renewcommand{\shortauthors}{Zimmermann et al.}

\ccsdesc[500]{Networks~Application layer protocols}
\ccsdesc[500]{Networks~Network measurement}

\keywords{HTTP/2, Server Push, Interleaving Push}

\begin{document}
\hypersetup{citecolor=,linkcolor=}
\begin{abstract}
HTTP/2 supersedes HTTP/1.1 to tackle the performance challenges of the modern Web. 
A highly anticipated feature is Server Push, enabling servers to send data without explicit client requests, thus potentially saving time.
Although guidelines on how to use Server Push emerged, measurements have shown that it can easily be used in a suboptimal way and hurt instead of improving performance.
We thus tackle the question if the current Web can make better use of Server Push.
First, we enable real-world websites to be replayed in a testbed to study the effects of different Server Push strategies.
Using this, we next revisit proposed guidelines to grasp their performance impact.
Finally, based on our results, we propose a novel strategy using an alternative server scheduler that enables to interleave resources.
This improves the visual progress for some websites, with minor modifications to the deployment.
Still, our results highlight the limits of Server Push: a deep understanding of web engineering is required to make optimal use of it, and not every site will benefit.
\end{abstract} \maketitle
\def\AlexaPushUsageAsAbs{{892}}

\def\AlexaPushPercentage{{0.09}}

\def\AlexaHTwoUsageAsAbs{{241,386}}

\def\AlexaHTwoPercentage{{24.14}}

\def\AlexaPushInHTwoPercentage{{0.37}}

\section{Introduction}

The Hypertext Transfer Protocol (HTTP) is the de-facto protocol for realizing desktop and mobile websites as well as applications.
Traffic shares of\,$>$\,\unit[50]{\%}, \eg in a residential access link~\cite{GregorIMC}, an IXP~\cite{SigcommIXP}, or backbone~\cite{SevenYears,rueth2018quic}, express this dominance.
Despite this, HTTP-based applications are built on top of a protocol designed nearly two decades ago, now suffering from various inefficiencies in the modern Web, \eg \ac{hol} blocking.
To address the drawbacks, HTTP/2 (\h{2}) was standardized~\cite{rfc7540} as \h{1}'s successor.
Among others, a highly anticipated feature~\cite{grigorik2013high} is \emph{Server Push}, changing the \emph{pull-only} into a \emph{push-enabled} Web.
It enables servers to send additional resources \emph{without} explicit requests, \eg send a CSS upon a request for \texttt{index.html}, thus saving round trips.

\begin{figure}[t]
	\begin{subfigure}[t]{0.5\columnwidth}
		\includegraphics{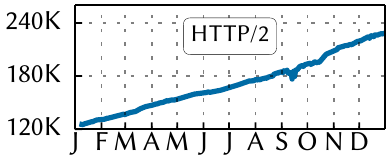}
		\label{intro:fig:deplyoment:h2}
	\end{subfigure}%
	\begin{subfigure}[t]{0.5\columnwidth}
		\includegraphics{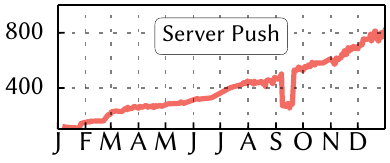}
		\label{intro:fig:deplyoment:push}
	\end{subfigure}%
	\vspace{-10pt}
	\caption{Adoption of HTTP/2 and Server Push over the course of one year\protect\footnotemark (2017) on the Alexa \unit[1]{M} list~\cite{zimmermann2017push}.
	Although the use of both is increasing, the adoption of Server Push is relatively low.}
	\label{intro:fig:deplyoment}
	\vspace{-10pt}
\end{figure}

This potential for speeding up the Web manifests in a growing interest in Server Push, both among CDNs~\cite{thumb,zafiris2017making} and in research~\cite{butkiewicz2015klotski,zimmermann2017push,kelton2017webgaze,ruamviboonsuk2017vroom}.
These efforts face the challenge that the standard only defines the Server Push protocol, not how to use it, \ie \emph{what} to push \emph{when}, determining its performance.
In this regard, previous work provided several strategies or approaches how to use Server Push.
They involve signaling clients what to fetch next~\cite{ruamviboonsuk2017vroom}, dependency analysis of content~\cite{butkiewicz2015klotski}, gaze tracking to identify regions of interest to be pushed~\cite{kelton2017webgaze}, or guidelines for its basic usage~\cite{thumb}.
However, its usage is still low relative to the \h{2} adoption (cf.~Fig.~\ref{intro:fig:deplyoment})~\cite{varvello2016web,zimmermann2017push}---potentially given its complex usage. 
In previous studies, we showed that Server Push can be easily used suboptimally in real-world deployments and \emph{hurt} instead of \emph{improving} performance~\cite{zimmermann2017push, zimmermann2017pushqoe}.
These findings, as well as discussions among web and protocol engineers~\cite{jake2017tough,pushissue,akamai2018ietf,chrome2018ietf}, highlight that the quest for \emph{optimal} Server Push usage is far from being settled.

\footnotetext{Live results available at \url{https://netray.io}}

We thus tackle the question \emph{what} influences the Server Push performance, \emph{how} push strategy performance can be reproducibly tested on any website, and \emph{how} a new strategy can speed up the current Web without major modifications.
Specifically,
\begin{inparaenum}[\em i)]%
we \item propose a new evaluation method to automatically and reproducibly evaluate Server Push strategies on any website by replay.
This complements measurements of real-world deployments and provides a method to systematically understand the isolated baseline performance of Server Push strategies for a broad class of websites.
We \item use this method to study the performance impact of strategies on real-world and synthetic websites.
In this regard, we revisit \emph{existing} strategies proposed in related work and find that pushing all objects, as a straightforward approach, can hurt the performance.
By varying the amount, order, and type of objects, we observe positive \emph{and} negative effects, highlighting the challenge of optimal usage.
Given these results and based on an analysis of the respective rendering process in the browser, we \item investigate a new approach to push the \emph{right resource} at the \emph{right time}, by interleaving the base document and pushed objects.
This \emph{can} lead to a faster visual progress for some popular websites. 
Still, our results also highlight the limits of Server Push, as it requires a deep understanding of the load and render process and not every website can benefit.
\end{inparaenum}

\section{Background}
\label{sec:bg}
We now briefly present an overview of \h{2}, as well as two prominent metrics to assess the performance of websites.

\subsection{H2 Overview}
\label{sec:bg:h2}
\h{2} was standardized in 2015~\cite{rfc7540} to replace \h{1}.
A key difference is a binary instead of an ASCII representation, allowing easier framing and parsing, which increases the processing efficiency~\cite{grigorik2013high}.
Further, \h{2} allows to \emph{multiplex} requests/responses over \emph{one} connection, leading to parallel \emph{streams}, identified by IDs.
Clients can \emph{prioritize} streams, \eg to prefer certain objects.
Multiplexing also reduces application layer \ac{hol} blocking, an issue in \h{1}.
Like \h{1}, \h{2} is \emph{stateless}, thus header information for a connection is repetitive.
As mitigation, \h{2} adds \emph{header compression}~\cite{rfc7541}.
Though not mandatory, most implementations use \h{2} via TLS.
Besides privacy and security benefits, a TLS-tunnel over deployed intermediaries eases the deployment of new protocols~\cite{grigorik2013high}.

\afblock{\h{2} Server Push.}
Finally, \h{2} adds \emph{Server Push}.
To grasp its potential benefits, recall the classic \h{1} request/response model: a browser requests the base document, parses it, and then requests all discovered  objects individually.
Contrary, an \h{2} server can \emph{push} objects without \emph{explicit} request, \eg a CSS upon a request to \texttt{index.html}, thus saving round trips.
Hence, \h{2} allows transferring resources \emph{before} the browser finishes parsing.
To push, a server announces information about the object and stream it will use.
Afterward, the data is sent. 
A server is only allowed to push content from origins under its authority.
Further, a client can cancel an announced push, which is useful if the object is already cached.
Also, a client can deactivate the feature by setting \texttt{SETTINGS\_ENABLE\_PUSH} in a specific settings header to \texttt{0} at connection startup.
Yet, as seen in own measurements, by the time a client cancels the push, the object can be already in flight.
Also, the standard does not include mechanisms to signal the cache status, but drafts and academic approaches exist~\cite{han2015meta,oku2017cache}.

\subsection{Website Performance Metrics}
\label{sec:bg:perf}

\afblock{PLT.} In recent years, the key metric to measure performance was, and still is, \acf{plt}~\cite{butkiewicz2015klotski,kelton2017webgaze,rosen2017push,varvello2016web,wang2014spdy,wang2013wprof}, which represents the time between events in the browser's W3C Navigation Timing API.
Mostly, it refers to the time of the \texttt{onload} event, \ie a resource and its dependent resources finished loading, but other events are used in related work as well.
Here, we define \ac{plt} as the time between the \texttt{connectEnd} event, \ie the connection is established (DNS, TCP and TLS) and start of the \texttt{onload} event.
Still, \ac{plt} can be an over- or underapproximation~\cite{kelton2017webgaze}, \eg events may refer to items not in view, and some resources can be loaded by scripts \emph{afterward}.
Hence, \ac{plt} can fail to capture human perception~\cite{bocchi2017web,zimmermann2017pushqoe}.

\afblock{SpeedIndex.} To overcome the limits of \ac{plt}, Google proposed SpeedIndex~\cite{speedindex} to capture the visual progress of \emph{above-the-fold} content, \ie content in the viewport without scrolling.
It expresses how \emph{complete} a website looks at various points of its loading process.
To calculate SpeedIndex, the loading process is recorded as a video and each frame is compared to the final frame, thus measuring completeness.
While this \emph{visual} metric is an improvement over \ac{plt}, capturing a video may only be feasible for studies in the lab but not in the wild~\cite{meenan2013fast,bocchi2017web}.
\section{Related Work}
\label{sec:rw}
Our work relates to approaches that focus on \h{2} and Server Push performance, as well as frameworks and guidelines.

\afblock{\h{2}.}
Wang~\etal{}~\cite{wang2014spdy} provided the first analysis of SPDY, \h{2}'s predecessor.
Comparing SPDY to \h{1} on the transport level, they observe benefits, especially for large objects and low loss, and for few small objects in good conditions with large TCP initial windows.
Also, Server Push \emph{can} improve \ac{plt} for large RTTs, but the analysis of several policies also reveals impairments.
De Saxc\'{e}~\etal~\cite{saxce2015h2} evaluate \h{2} and focus on latency and loss, showing that \h{2} is less prone to higher latencies than \h{1}.
They regard Server Push as valuable, but more performance research is required.
Varvello~\etal~\cite{varvello2016web} present an adoption study (Alexa \unit[1]{M}) and compare the real-world performance of \h{2} websites to their \h{1} counterparts.
They observe benefits for \unit[80]{\%} of websites, but also degradations, without an explicit focus on Server Push.

\afblock{Server Push.}
In previous work~\cite{zimmermann2017push}, we complement the view on \h{2}'s adoption provided in~\cite{varvello2016web} and target broader sets, \ie IPv4 scans and all .com/.net/.org domains, and Server Push explicitly.
While the adoption of \h{2} \emph{and} Server Push is rising, the latter is orders of magnitude lower (cf.~Fig.~\ref{intro:fig:deplyoment}).
Using different protocol settings, we observe that Server Push can improve as well as hurt the performance, but the results cannot be mapped to simple reasons (\eg amount of bytes pushed), and that further analysis is required.
Instead of resources, Han~\etal~\cite{han2015meta} propose to push \emph{hints}, enabling the client to request critical resources earlier, which \emph{can} improve the performance.
Rosen~\etal{}~\cite{rosen2017push} analyze the benefits and challenges of Server Push and show that network characteristics play a major role in the effectiveness, similar to~\cite{wang2014spdy, saxce2015h2}.
As one guideline, they propose to push as much as possible, which not \emph{always} leads to improvements.
Butkiewicz~\etal~\cite{butkiewicz2015klotski} present Klotski, which prioritizes high-utility resources, \eg above-the-fold or by user preferences, obtained via surveys, live ratings, and offline measurements.
To deliver those, Server Push is used.
Kelton~\etal~\cite{kelton2017webgaze} prioritize objects of visual interest, identified in user studies via gaze tracking.
As objects can depend on each other, they employ dependency analysis~\cite{wang2013wprof} to prioritize objects and send high priority objects via Server Push.
Still, as there are impairments for some websites, they revert to the default setting in such cases. 
Vroom~\cite{ruamviboonsuk2017vroom} uses a client scheduler that parses \emph{preload} headers~\cite{grigorik2017preload}, containing dependency hints for resources (on other servers) to be fetched. 
As in~\cite{butkiewicz2015klotski}, high priority resources are pushed.
This combination can improve performance compared to base \h{2}.
In parallel to our work, results for the real-world performance of Server Push presented at the IETF \cite{akamai2018ietf, chrome2018ietf} indicated that more measurements are needed to grasp its benefits and even provoked the discussion, if Server Push should be focused on in the future and be used at all.

\afblock{Rules of Thumb for Push.}
Bergan~\etal~\cite{thumb}~provide a set of rules, evaluated in different simulations, network settings, and websites.
We focus on rules relevant for this paper and refer to~\cite{thumb} for more information.
\begin{inparaenum}[\em i)]%
A server should 
\item push just enough to fill idle network time.
As soon as the browser parses the HTML and requests objects, using push is not faster.
This is contrary to~\cite{rosen2017push}, suggesting to push as much as possible.
Further, objects should \item be pushed in evaluation-dependent order, as suboptimal orders can \eg delay the discovery of hidden resources loaded by scripts.
\end{inparaenum}

\afblock{\textit{Takeaway:~}}
Despite existing frameworks and guidelines, \h{2} Server Push is still not widely in use, and if, improvements are not guaranteed, as shown by recent measurements of real-world deployments.
Our goal is to provide an approach to understand the isolated performance of Server Push systematically and to shed light on its applicability for the current Web, \ie \emph{without} major modifications.
\section{Replaying Push Websites}
\label{sec:replay}

Although assessing the performance of Server Push in real-world deployments~\cite{varvello2016web, zimmermann2017push, akamai2018ietf, chrome2018ietf} is crucial, it imposes practical challenges. 
\begin{inparaenum}[\em i)]%
Among other factors, websites
	\item can change due to dynamic third-party content, \eg ads~\cite{goel2017measuring}, or  
	\item are subject to varying network characteristics.
\end{inparaenum}
This can cause misinterpretations of the results.
Hence, our goal is to understand the \emph{isolated} performance of Server Push under deterministic conditions, to reduce variability, and to enable reproducibility.
Therefore, we use a testbed to replay real-world websites and to grasp the possible potential of Server Push.
We exemplify this by using \textit{different} strategies and revisit existing guidelines, \eg~\cite{rosen2017push, thumb}, to assess their overall impact.
Next, we present our testbed and evaluation of different strategies.

\subsection{Controlled Push Strategy Evaluation}
\label{sec:replay:testbed}

We base our testbed, which we make publicly available together with more results~\cite{testbed}, on \textsc{Mahimahi}~\cite{netravali2015mahimahi}.
This enables to record \h{1} traffic to a database as request/response pairs, \eg captured in a browsing session.
Later, this database matches requests to responses to replay websites.
Network namespaces are used to recreate the deployment, \ie for each IP a local server is spawned. 
Thus, the same connection pattern as in the Internet is used, opposed to less realistic setups, such as local mirroring (\eg HTTrack) or proxies.

However, at the time of writing, the current version of \textsc{Mahimahi} does not support \h{2}.
To enable this support in \textsc{Mahimahi}, we first use the \h{2}-capable \texttt{mitmproxy}~\cite{mitmproxy} to capture request/responses and convert it to \textsc{Mahimahi}'s record format. 
We add the \texttt{h2o} webserver~\cite{h2o} and create an \texttt{h2o}-FastCGI module that matches and serves responses from the record DB. 
Finally, we enable to specify \emph{push strategies}, to define responses to be pushed.
\h{2} supports connection-coalescing, \ie a server with a single IP can be authoritative for multiple domains and serve content via \emph{one} connection.
Currently, a browser handles traffic for different origins over the \emph{same} connection if a server presents a TLS certificate that includes the origins as \emph{Subject Alternative Names}. 
The browser also checks if IPs of different origins match using DNS.
We thus modify \textsc{Mahimahi} to generate certificates for \emph{each} local server, which include domains with the same IP.
Also, we assume \emph{every} server to be \h{2}-enabled, as in ~\cite{ruamviboonsuk2017vroom}.
Using \texttt{tc}, we simulate DSL settings with \unit[50]{ms} RTT and \unit[16]{Mbit/s} down- and \unit[1]{Mbit/s} uplink between the client and the servers.
Please note that we do not assume any additional delay \emph{on} the servers, \eg for additional fetches or disk access.
For all evaluation settings, we follow settings used in previous work~\cite{zimmermann2017pushqoe} and utilize \texttt{browsertime}~\cite{browsertime} to automate Chromium 64 and replay each website in each setting for 31 times.
If not stated otherwise, we show and discuss the median result of these repeated runs.

\begin{figure}[t!]
\begin{subfigure}[t]{\columnwidth}
	\includegraphics[width=\columnwidth]{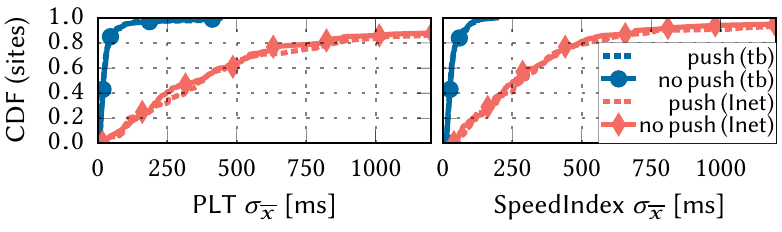}
	\vspace{-15pt}
	\subcaption{Comparing the std. error $\sigma_{\overline{x}}$ of the performance of 100 websites in our testbed (tb) vs. the Internet (Inet) over 31 runs.}
	\label{fig:testbed:evaluation:error}
\end{subfigure}
\begin{subfigure}[t]{\columnwidth}
	\includegraphics[width=1.0\columnwidth]{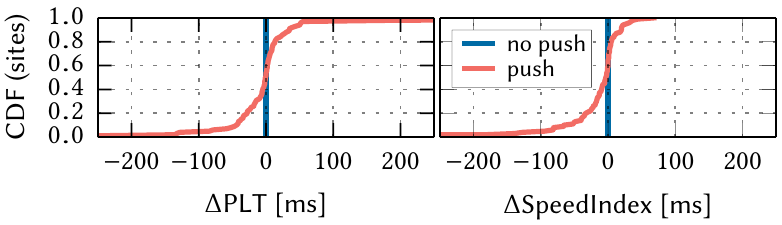}
	\vspace{-15pt}
	\subcaption{In our testbed, we observe similar effects as in the Internet, \ie improvements and detriments with Server Push when compared to the \emph{no push} case ($\Delta$\,\textless\,0 is better).}
	\label{fig:testbed:evaluation:change}
\end{subfigure}
\vspace{-10pt}
\caption{Evaluation of our realized testbed based on \textsc{Mahimahi}.}
\label{fig:testbed:evaluation}
\vspace{-17.5pt}
\end{figure}

\afblock{Evaluation.}
We evaluate our testbed by replaying 100 random websites using Server Push from Alexa \unit[1]{M}, with and without Server Push.
Fig.~\ref{fig:testbed:evaluation:error} shows the standard error for \ac{plt} and SpeedIndex (cf. Sec.~\ref{sec:bg:perf}), compared to their deployment in the Internet.
For \unit[95]{\%} (\unit[85]{\%}) of websites, the error $\sigma_{\overline{x}}$ is\,\textless\,\unit[100]{ms} (\unit[50]{ms}) for \ac{plt}, with similar results for SpeedIndex.
In the Internet, this holds only for \unit[14]{\%} (\unit[5]{\%}), which we attribute to varying network conditions.
These results indicate that we remove a lot of variability for the performance results of most websites, which helps to assess the performance impact in a reproducible manner.
To evaluate if we still, after removing this variability, observe performance improvements and impairments as the Internet~\cite{zimmermann2017push}, we compare the performance of Server Push against a \emph{no push} configuration, \ie the client signals the server to disable Server Push (cf.~Sec.~\ref{sec:bg:h2}).
Pushing the \emph{same} objects as in the Internet, we observe no benefits for \unit[49]{\%} (\unit[35]{\%}) of websites in PLT (SpeedIndex) (cf.~Fig.~\ref{fig:testbed:evaluation:change}).
We thus argue that our testbed enables to reproduce results for a lot of websites and we still observe positive \emph{and} negative effects (cf.~Sec.~\ref{sec:rw}).

\subsection{Real-World: Altering What to Push}
\label{sec:real_world}
Next, we replay real-world websites subject to various push strategies in our testbed, which enables us to study the impact on performance in a \emph{controlled} environment.
To this end, we create two disjunct random sets of 100 websites (HTTPS) each, one from the top 500 (top-100) and from the top \unit[1]{M} (random-100) according to Alexa.
We use the Alexa list as a basis, as we expect a lot of \h{2}-enabled websites among this popular list~\cite{zimmermann2017push, scheitle2018toplist}.
If there is \emph{no} \h{2} version, we capture the respective \h{1} version. 
Please note that in case of an \h{2}-enabled website, we do not capture if the website uses Server Push, as our goal is to apply \emph{different} strategies.

\begin{figure*}[t]
\begin{subfigure}[t]{\columnwidth}
	\includegraphics[width=\columnwidth]{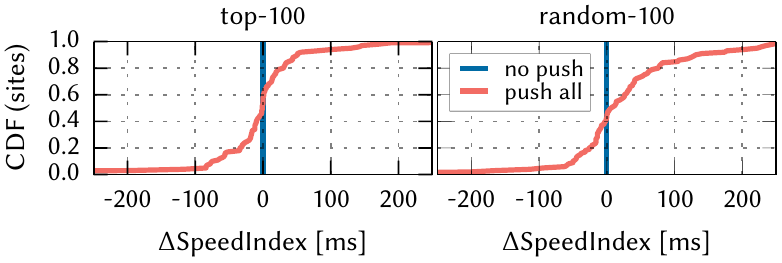}
	\subcaption{SpeedIndex when pushing all objects (in request order) normed to the \emph{no push} case. \ac{plt} shows a similar behavior.}
	\label{fig:testbed:push_all}
\end{subfigure}
\begin{subfigure}[t]{\columnwidth}
	\includegraphics[width=\columnwidth]{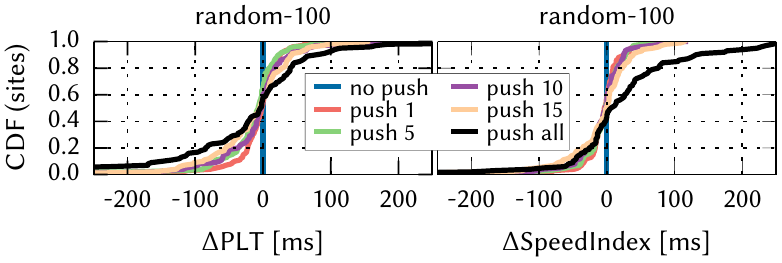}
	\subcaption{Push limited amount, normed to the \emph{no push} case. Only for the random set, due to limited objects for sites in the top set.}
	\label{fig:testbed:alter_push:amount}
\end{subfigure}
\vspace{-10pt}
\caption{Delta (median) when pushing all (Fig.~\ref{fig:testbed:push_all}) and a limited amount of objects (Fig.~\ref{fig:testbed:alter_push:amount}), using a computed order.}
\label{fig:testbed}
\vspace{-10pt}
\end{figure*}

\afblock{Pushable Objects.}
For the top-100 (random-100) set, \unit[52]{\%} (\unit[24]{\%}) have \textless\,\unit[20]{\%} of pushable objects, \ie the other objects reside on servers beyond the authority of the pushing server. 
Hence, many websites \emph{cannot} push all objects. 

\afblock{Computing the Push Order.}
Next, we access the websites in our testbed via \h{2} 31 times, not pushing \emph{any} objects.
We trace requests and priorities (cf.~Sec.~\ref{sec:bg}) used to obtain the landing page and construct a dependency tree, based on these priorities. 
By traversing this tree, we compute a request order.
The goal is to obtain the desired order used by the browser to request objects, to identify an order to push.
Here, we follow the rules given in~\cite{thumb} (cf.~Sec.~\ref{sec:rw}), as suboptimal orders can have negative impacts, \eg delay critical resources.
Since the order is not stable across all runs, \eg due to client-side processing, we use a majority vote.
As this order is based on the \emph{initial} connection to the origin server, all objects are pushable (cf. Sec.~\ref{sec:bg:h2}).

\subsubsection{Varying Amount and Type of Objects}
\label{sec:real_world:amount_type}
First, we push all \emph{pushable} objects following the computed order, as \emph{push all} is considered a valuable strategy in~\cite{rosen2017push}.
Fig.~\ref{fig:testbed:push_all} shows the results for SpeedIndex.
In the top-100 (random-100) set only \unit[58]{\%} (\unit[45]{\%}) of sites benefit (\ac{plt} similar but not shown).
Pushing \emph{all} objects can be harmful, as it can delay processing, and even in the positive case, waste bandwidth or cause contention on the server~\cite{han2015meta,ruamviboonsuk2017vroom}.
Thus, we vary the amount of objects $n \in \{1,5,10,15\}$ to push.
Here, we only consider the random-100 set, as not all top-100 websites have enough pushable resources.
We again use fixed orders, limited to the first $n$ objects.
For a small $n$, this is in line with guidelines~\cite{thumb} (cf.~Sec.~\ref{sec:rw}) to push just enough to fill the network idle time.
We observe (cf.~Fig.~\ref{fig:testbed:alter_push:amount}) that pushing less can lead to less detrimental effects, \eg delay of processing and requests.
Still, a lot of websites exhibit no \emph{significant} improvements.

Last, we analyze the impact of pushing specific object types, again for the random-100 set.
While pushing CSS or JavaScripts (JS) can lead to positive and negative effects (no figure shown), images lead to a \emph{worse} SpeedIndex for \unit[74]{\%} of websites.
This is no surprise, as they do not contribute to the creation of the Document- (\acs{dom}) or \ac{cssom}---both essential parts of the layout and render process.
Using the \emph{best} type strategy per website, \ie if pushing CSS is better than JS, CSS is the best type strategy, only \unit[24]{\%} (SpeedIndex) and \unit[20]{\%} (PLT) of websites improve.
Type combinations, \ie CSS+JS and CSS+images, lead to similar results.

Also, we analyzed the effect if we vary the computed push order (results not shown).
We observe that the impact on performance is highly dependent on the overall amount of objects, the structure of the HTML, \ie when objects are referenced, and object type.
Exemplary, a suboptimal order could prefer uncritical resources, with respect to above-the-fold, and thus delay critical resources.  

\afblock{\emph{Conclusion:}} We observe that a large fraction of resources is not pushable as they reside on other servers. 
Moreover, many websites depend on third-party content---impacting the loading process~\cite{butkiewicz2011complexity,goel2017measuring}.
Also, not many sites benefit from a \emph{push all} strategy.
Pushing \emph{less} can reduce negative effects, but not always improves respective metrics. 
In addition, websites \emph{can} benefit from different object types to be pushed.
Overall, we do not find an automatically generated \emph{one-fits-all} strategy.

\subsection{Synthetic Sites and Custom Strategies}
\label{sec:synthetic}
\begin{figure}[t]
	\centering
	\includegraphics[width=1.0\columnwidth]{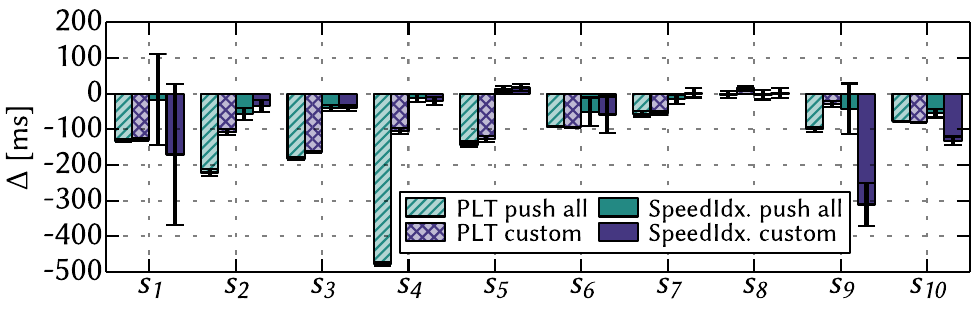}
	\vspace{-22.5pt}
	\caption{Custom strategies normed to the \emph{no push} case. We show the average and the \unit[95]{\%} confidence interval ($\Delta$\,\textless\,0 is better).}
	\label{fig:testbed:synthetic}
	\vspace{-15pt}
\end{figure}

Up to now, we assumed real-world settings, \eg content at several servers, which can have unpredictable performance impacts.
Thus, we use \emph{synthetic} websites \emph{s\textsubscript{1}}-\emph{s\textsubscript{10}} that are snapshots of websites or templates, and deploy them on a \emph{single} server, \ie we relocate content.
Again, we use request orders for \emph{push all} and \emph{no push} as a baseline.
Instead of automatic generation (cf.~Sec.~\ref{sec:real_world}), we create \emph{custom} strategies.
We inspect the browser's loading and rendering process and select resources that
\begin{inparaenum}[\em i)]%
	\item appear above-the-fold, or
	\item are required to paint above-the-fold content.
\end{inparaenum}
This can lead to benefits (cf.~Fig.~\ref{fig:testbed:synthetic}), which we discuss in two case studies.
We focus on time and size improvements, as pushing less is preferable (cf.~Sec.~\ref{sec:real_world:amount_type}).

\afblock{Case Studies.}
\emph{s\textsubscript{1}} shows a loading icon that fades and content is shown once the DOM is ready.
Thus, we push resources blocking the DOM construction (JS, CSS) and \emph{hidden} fonts referenced in the CSS.
On average, this improves SpeedIndex (but fluctuates), by only pushing \unit[309]{KB} (\unit[1,057]{KB}, \emph{push all}).

\emph{s\textsubscript{5}} takes \unit[692]{ms} (\emph{push all})~vs.~\unit[1,038]{ms} (\emph{no push}) to be transferred, but the metrics do not significantly improve.
Regarding \ac{plt}, a blocking JS is referenced late in the \texttt{<body>}, which requires to create the CSSOM.
This takes longer than the transfer, and the browser is not network but \emph{computation} bound, affecting the overall process, similar to results in~\cite{ruamviboonsuk2017vroom}.
Our strategy pushes four render-critical resources and some images.
Yet, there is no benefit, as the browser can request resources as fast as the server could push them, due to a large HTML, resulting in no network idle time.
We notice the same for \emph{s\textsubscript{8}}.
The HTML transfer requires multiple round trips to be completed.
After the first chunk, the browser can issue requests for six render-critical resources referenced early, and using push does not change the result, similar to findings in~\cite{thumb}.

\afblock{\emph{Conclusion:}} Pushing all resources in a request order in this setting, \ie \emph{all} resources hosted on a single server, \emph{can} reduce \ac{plt} compared to \emph{no push}, but SpeedIndex rarely improves. 
Moreover, we do not observe significant detrimental effects.
Yet, pushing everything can be wasteful in terms of bandwidth, \eg if the resource is already cached, and cause contention between objects~\cite{kelton2017webgaze}.
For some websites, our custom strategy performs equally to \emph{push all} by pushing \emph{fewer} resources.
Still, even by manual inspection of the page load process, we are unable to optimize the SpeedIndex significantly for many websites.
We thus conclude that the \emph{optimal} push strategy is \emph{highly} website-specific and requires manual effort, an in-depth understanding of the page load and render process, as well as the interplay of resources. %
\section{Interleaving Push}
\label{sec:interleaving}

\begin{figure}[t]
\captionsetup[subfigure]{justification=justified}
\begin{subfigure}[t]{\columnwidth}
	\includegraphics[width=\columnwidth]{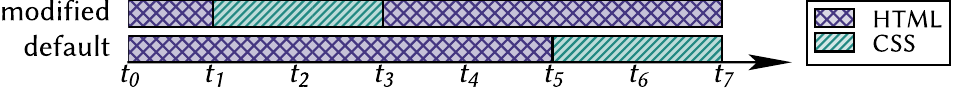}
	\subcaption{\texttt{h2o}'s default scheduler treats a push (\eg CSS) as child of the parent stream (HTML).
	If the parent does not block, the \emph{entire} stream is sent, possibly delaying critical resources.
	Our modification stops after a defined offset to start pushing.}
	\label{fig:interleaving:scheduler}
	\vspace{5pt}
\end{subfigure}
\captionsetup[subfigure]{justification=centering}
\begin{subfigure}[t]{\columnwidth}
	\includegraphics[width=1.0\columnwidth]{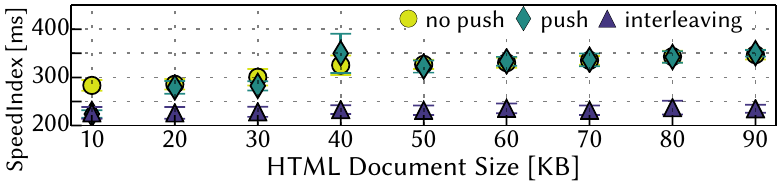}
	\subcaption{SpeedIndex (test website) for different strategies. Our \emph{interleaving} strategy yields a stable time (average and std. dev.).}
	\label{fig:interleaving:motivation}
\end{subfigure}
\vspace{-11pt}
\caption{Interleaving Push concept and performance example.}
\label{fig:interleaving}
\vspace{-20pt}
\end{figure}

We saw that the benefits of push depend on the size of the base document and the position of resource references, \eg pushing objects referenced late in large base documents can be beneficial while pushing early referenced objects may not.
We also observed that the order of pushes is performance critical~\cite{thumb}~(cf.~Sec.~\ref{sec:real_world}).
Thus, our goal is to analyze if \emph{interleaving} the base document with pushed objects can be beneficial.
The intuition is to push the \emph{right resources} at the \emph{right time}.

\afblock{Motivating Example.}
We create a website that references CSS in the \texttt{<head>} section and vary the size of the \texttt{<body>} by adding text.
As baseline,
\begin{inparaenum}[\em i)]%
	the \item browser requests the CSS \emph{(no push)}. Next, we 
	\item \emph{push} the CSS upon request for the HTML. Last, we
	\item \emph{interleave} the delivery of HTML such that after a fixed offset, the server makes a \emph{hard} switch to push the CSS, before proceeding with the HTML.
\end{inparaenum}
Thereby, we incorporate page-specific knowledge into the strategy. 

For the latter, we modify \texttt{h2o}'s stream scheduler.
Per default, a push is treated as child of the parent stream, \eg a CSS as child of the \texttt{index.html}.
Thus, the server pushes if the parent stream blocks, \eg due to extra fetches if not present at the server, or is finished.
Our \emph{modification} stops the parent stream after a \emph{defined} offset, \eg after \texttt{</head>} and first bytes of \texttt{<body>}, and starts to push (cf.~Fig.~\ref{fig:interleaving:scheduler}).
In the \emph{no push} case, Chromium assigns a lower priority to CSS than for the HTML.
The \texttt{h2o} server adheres to this and sends the CSS \emph{after} the HTML.
Here, \emph{no push} and \emph{push} perform similar (cf.~Fig.~\ref{fig:interleaving:motivation}), as the parent does not block.
\emph{Interleaving} via push yields nearly constant and faster performance.

\begin{figure*}[t]
\begin{subfigure}[t]{\columnwidth}
	\includegraphics[width=\columnwidth]{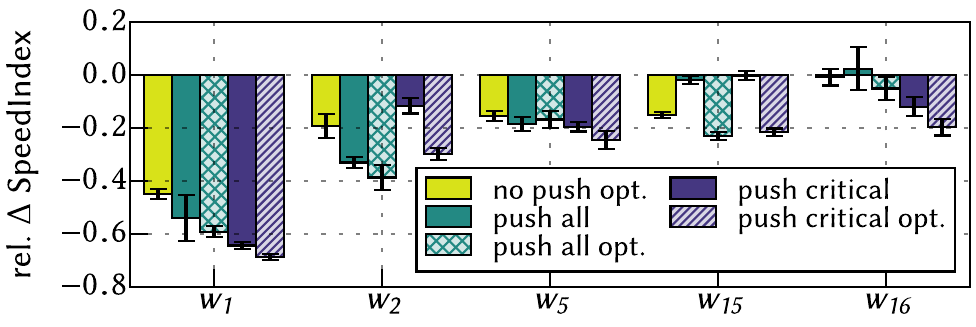}
	\vspace{-15pt}
	\subcaption{Websites with improvements ($\ge$\,\unit[20]{\%}) for our \emph{push critical optimized} strategy over the \emph{no push} case.}
	\label{fig:interleaving:evaluation:good}
\end{subfigure}
\begin{subfigure}[t]{\columnwidth}
	\includegraphics[width=\columnwidth]{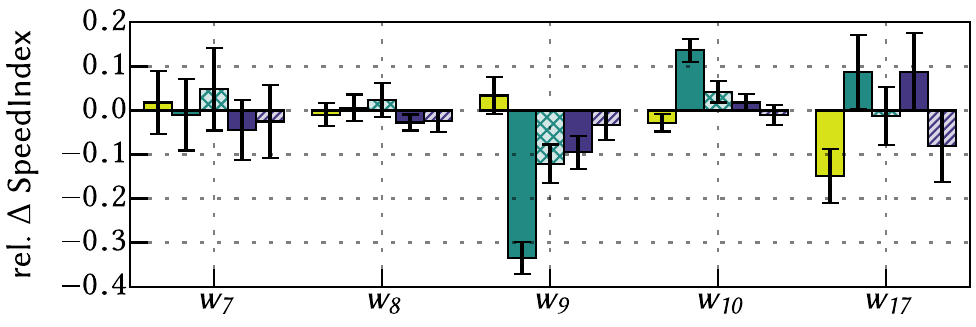}
	\vspace{-15pt}
	\subcaption{Exemplary websites with detrimental effects or no benefits (\textless\,\unit[10]{\%}) for \emph{push critical optimized} vs. \emph{no push}.}
	\label{fig:interleaving:evaluation:bad}
\end{subfigure}
\vspace{-10pt}
\caption{Performance of strategies (websites as in Tab.~\ref{tab:interleaving:selection}). We show avg. relative changes ($\Delta$\,\textless\,0 is better) and 99.5\,\% confidence.}
\label{fig:interleaving:evaluation}
\vspace{-10pt}
\end{figure*}

\afblock{Real-World Websites.} 
Motivated by this potential, we now focus on the applicability on real websites.
Our set \emph{w\textsubscript{1}}-\emph{w\textsubscript{20}} (cf.~Tab.~\ref{tab:interleaving:selection}) covers a broad range of content \emph{and} website structures, \eg \emph{w\textsubscript{5}} consists of 8 requests served by one server, while \emph{w\textsubscript{17}} consists of 369 requests to 81 servers.
Given this complexity and our prior analysis of browser internals, we perform measurements with \emph{distinct modifications}.
We unify domains of the same infrastructure, \eg \texttt{img.bbystatic.com} and \texttt{bestbuy.com}, 
and, based on inspection of the rendering process, host critical above-the-fold resources.

\begin{table}[t]
	\setlength{\aboverulesep}{0pt}
	\setlength{\belowrulesep}{0pt}
	\small
	\setlength{\tabcolsep}{3.45pt}
	\begin{tabularx}{\columnwidth}{ll ll ll ll}
		\toprule
		\rowcolor{gray!25}\emph{w\textsubscript{1}}	& wikipedia*	&\emph{w\textsubscript{6}}	&	chase			&\emph{w\textsubscript{11}}&aliexpress		& \emph{w\textsubscript{16}}	& twitter\textsuperscript{\textdagger}\\
		\emph{w\textsubscript{2}}					& apple		&\emph{w\textsubscript{7}}	&	reddit			&\emph{w\textsubscript{12}}&ebay			& \emph{w\textsubscript{17}}	& cnn\\					
		\rowcolor{gray!25}\emph{w\textsubscript{3}}	& yahoo		&\emph{w\textsubscript{8}}	&	bestbuy			&\emph{w\textsubscript{13}}&yelp			& \emph{w\textsubscript{18}}	& wellsfargo\\
		\emph{w\textsubscript{4}}					& amazon		&\emph{w\textsubscript{9}}	&	paypal			&\emph{w\textsubscript{14}}&youtube		& \emph{w\textsubscript{19}}	& bankofamerica\\
		\rowcolor{gray!25}\emph{w\textsubscript{5}}	& craigslist	&\emph{w\textsubscript{10}}	&	walmart			&\emph{w\textsubscript{15}}&microsoft		& \emph{w\textsubscript{20}}	& nytimes\\
		\Xhline{0.75pt}
	\end{tabularx}
	\caption{Websites (.com) selected for \emph{interleaving push}.
	If not marked, we capture the landing page. (*Article, \textsuperscript{\textdagger}Profile.)}	
	\label{tab:interleaving:selection}
	\vspace{-30pt}
\end{table}
 
\afblock{Strategies.} 
As the baseline, we
\begin{inparaenum}[\em i)]%
	\item use a \emph{no push} strategy. We extend this to 
	\item a \emph{no push optimized} strategy, where we use \texttt{penthouse}~\cite{penthouse} to compute a \emph{critical CSS} from the included CSS, that is required to display above-the-fold content, inspired by~\cite{butkiewicz2015klotski,thumb}.
We reference this critical CSS in the \texttt{<head>} section and all \emph{other} CSS at the end of the \texttt{<body>}. In
	the \item \emph{push all} strategy, we push \emph{all} resources hosted on the previously merged domains, which might include additional non-critical resources.
We extend this setting to the 
	\item \emph{push all optimized} strategy.
Here, we first push the critical CSS and critical above-the-fold resources in an interleaved fashion, and after the HTML, all other pushable resources.
In the
	\item \emph{push critical} strategy, we push \emph{only} critical resources for above-the-fold content.
Last, \item the \emph{push critical optimized} strategy adds the critical CSS modification to the prior strategy.
\end{inparaenum}
For all \emph{optimized} strategies, we use the \emph{modified} server and the \emph{default} in all other cases.

\afblock{Evaluation.} Using the \emph{push critical optimized} strategy, we see benefits for five websites (cf. Fig.~\ref{fig:interleaving:evaluation:good}), but impairments or no advance (\textless\,10\,\%) for all others.
Next, we focus on a representative set (cf. Fig.~\ref{fig:interleaving:evaluation:bad}), discuss influence factors, and summarize\footnote{{More results available at \url{https://push.netray.io}}}.
Reported changes are averages and sizes are obtained on the protocol level.

The SpeedIndex of \emph{w\textsubscript{1}} is reduced by \unit[44.95]{\%} using only a critical CSS (\emph{no push optimized}).
With the \emph{push all optimized} strategy, we see an improvement of \unit[59.19]{\%}, and even \unit[68.85]{\%} for \emph{push critical optimized}.
In the latter, we push \unit[78.43]{KB} compared to \unit[1,123]{KB}, saving \unit[93]{\%}.
Here, interleaving push is beneficial, because of a large HTML size (\unit[236]{KB} compressed).
In the \emph{no push} case, the browser prioritizes the HTML over the CSS, and thus the server first sends the \emph{entire} HTML.
In our case, we push critical CSS after \unit[4]{KB} of HTML, enabling to construct the DOM faster, and also push a blocking JS and two images, before continuing with the HTML.

\emph{w\textsubscript{2}} already shows an improvement of \unit[19.22]{\%} when using a critical CSS in comparison to the \emph{no push} strategy.
In combination with the \emph{push all} and \emph{push critical} strategy, this yields the best performance, \ie improvements of \unit[33.05]{\%} and \unit[38.7]{\%}.
Our \emph{push critical optimized} strategy achieves a competitive improvement of \unit[29.74]{\%}, only pushing \unit[289.63]{KB} instead of \unit[725.57]{KB} compared to the \emph{push all optimized} strategy, saving $\sim$\,\unit[60]{\%}.
In the default case, several CSS requested \emph{after} the HTML block the execution of JS and thus the DOM construction.
With interleaving push, we extract the critical CSS and thus reduce the critical render path.

For \emph{w\textsubscript{16}}, creating a critical CSS is not beneficial, as the website already uses such optimizations.
Still, using our \emph{push critical optimized} strategy improves performance by \unit[19.67]{\%}, pushing \unit[10.2]{KB} of resources.
\emph{w\textsubscript{16}} has a similar setting as \emph{w\textsubscript{1}}, \ie CSS is made dependent on the HTML (\unit[45]{KB} compressed).
With interleaving push, the server starts pushing CSS after \unit[12]{KB} of HTML, allowing to proceed with the construction of the DOM faster. 

However, other websites exhibit no major improvements or even detrimental effects.
Though we are able to remove \unit[87]{KB} from the critical render path of website \emph{w\textsubscript{7}} by pushing the critical CSS instead of all CSS, the overall visual progress is not affected as much, because \emph{w\textsubscript{7}} contains a large blocking JS in the \texttt{<head>}. 
Similar effects are observed for \emph{w\textsubscript{8}}.

\emph{w\textsubscript{9}} benefits from pushing all resources.
Still, a critical CSS does not yield drastic improvements, as the HTML contains no blocking code until the end, \ie no delay of processing.

For \emph{w\textsubscript{10}}, we see detrimental effects by pushing all resources, as the page contains a lot of images, which causes bandwidth contention with other push streams.
Pushing only critical resources reduces detrimental effects, but does not improve over  \emph{no push}. 
We find that a large portion of JS is inlined into the HTML.
Therefore, interleaving push is not as efficient.

\emph{w\textsubscript{17}} benefits from a critical CSS in the \emph{no push optimized} case, improving by \unit[14.88]{\%}, but using push does not yield improvements \textgreater\,\unit[8]{\%}.
By manual inspection, we see that pushing improves the time of the \emph{first} visual change, but not the SpeedIndex, which we attribute to the structural complexity and amount of requests.

\afblock{Summary.} We evaluated a novel way to utilize Server Push.
By interleaving the HTML with critical CSS and critical above-the-fold resources on the \h{2} frame level, we can improve some websites in our testbed.
Still, we also see that the benefits \emph{highly} depend on the underlying website's structure and have to be evaluated individually.
This requires a \emph{deep} understanding of the page load and rendering process in the browser.
Most promising examples include websites where we find critical blocking resources, \ie CSS or JS affecting the DOM construction, referenced early.
We also observe that switching to pushing critical resources while the browser processes inlined JS can also be beneficial, similar to~\cite{ruamviboonsuk2017vroom}.

Still, many websites do not benefit from our optimizations in our testbed, based on various reasons.
Some websites already employ optimizations such as inlining critical JS or CSS, such that a browser is not blocked after receiving the first bytes of HTML, limiting the effect of interleaving push.
Also, we see that if a website contains a lot of third-party resources, \eg \emph{w\textsubscript{17}}, the effects of interleaving push dilute due to the complexity of the entire page load process. %
\section{Discussion}
We observed that the \emph{optimal} push strategy is \emph{highly} website specific and requires in-depth analysis of the page load process.
In the following, based on our findings in Sec.~\ref{sec:interleaving}, we discuss how a CDN \emph{could} employ our testbed and interleaving push approach as one possibility to generate strategies automatically.

\afblock{Use in CDN Deployments.} Some CDNs already employ \emph{Real User Measurements} (RUM) to obtain browser feedback, \ie embedded JavaScripts report the data obtained in the client's browser (\eg resource timings or rendering events) back to the CDN for further analytics~\cite{akamai2018ietf}.
Based on information about critical resources and rendering, several (interleaving) push strategies for different versions of a website and network settings, \eg mobile, desktop, cable or cellular, could be analyzed in our testbed.
Subsequently, the performance of these strategies could be assessed in A/B tests in a real deployment against the original version~\cite{fastly2016ab,chrome2018ietf,cloudflare2018ab,akamai2018ietf}. 
By incorporating this feedback, we believe it could be possible to learn website and browser-specific push strategies from browser interactions with CDN edge nodes.
However, testing the feasibility of this approach is beyond the scope of this paper.

\section{Conclusion}
This paper investigates if the current Web is ready for Server Push.
To systematically answer this question, we create an \h{2} testbed based on \textsc{Mahimahi}, which we open sourced~\cite{testbed}, to replay \emph{any} website in a controlled manner and subject to \emph{any} Server Push strategy.
We thoroughly study the influence of various Server Push strategies, \ie \emph{automatically} generated or based on \emph{guidelines}, on two major performance metrics, \ie \ac{plt} and SpeedIndex, for both a set of real-world and synthetic websites.
Our results indicate that a recommended strategy to push all embedded objects can optimize the performance of some sites, but also decrease the performance for others.
By further varying the amount, order, and type of pushed objects, we again observe benefits and detrimental effects as well, highlighting the fundamental challenge of optimal Server Push usage.
By \emph{tailoring} custom strategies and using a novel resource scheduler, we show that the performance of some popular sites can indeed be improved, with minor modifications to the deployment.

We find that, while the Web may be technically ready to support Server Push, it is no feature that can be utilized easily.
If and how Server Push should be used is subject to a number of \emph{website-specific} aspects.
Non-site specific adoption can very easily lower the web performance.
Thus, no general guidelines can be provided for optimal push usage, making the feature not straightforward to apply.
The question here is not if the Web is ready for Server Push but if the web engineers are eager to manual tuning.

\section*{Acknowledgements}
We thank our shepherd Ramesh Sitaraman and the anonymous reviewers for their insightful comments and suggestions.
This work has been funded by the DFG as part of the CRC 1053 MAKI. 
\balance
\bibliographystyle{ACM-Reference-Format}
\bibliography{literature}


\begin{thebibliography}{40}

%%% the Web address.

\ifx \showCODEN    \undefined \def \showCODEN     #1{\unskip}     \fi
\ifx \showDOI      \undefined \def \showDOI       #1{#1}\fi
\ifx \showISBNx    \undefined \def \showISBNx     #1{\unskip}     \fi
\ifx \showISBNxiii \undefined \def \showISBNxiii  #1{\unskip}     \fi
\ifx \showISSN     \undefined \def \showISSN      #1{\unskip}     \fi
\ifx \showLCCN     \undefined \def \showLCCN      #1{\unskip}     \fi
\ifx \shownote     \undefined \def \shownote      #1{#1}          \fi
\ifx \showarticletitle \undefined \def \showarticletitle #1{#1}   \fi
\ifx \showURL      \undefined \def \showURL       {\relax}        \fi
% invisible to TeX
\providecommand\bibfield[2]{#2}
\providecommand\bibinfo[2]{#2}
\providecommand\natexlab[1]{#1}
\providecommand\showeprint[2][]{arXiv:#2}

\bibitem[\protect\citeauthoryear{??}{bro}{}]%
        {browsertime}

\newblock \bibinfo{title}{{\texttt{browsertime}}}.
\newblock
  \bibinfo{howpublished}{\url{https://github.com/sitespeedio/browsertime}}.
\newblock
\newblock
\shownote{Online 06/18/2017.}


\bibitem[\protect\citeauthoryear{??}{h2o}{}]%
        {h2o}

\newblock \bibinfo{title}{{\texttt{h2o}}}.
\newblock \bibinfo{howpublished}{\url{https://h2o.examp1e.net}}.
\newblock
\newblock
\shownote{Online 06/18/2017.}


\bibitem[\protect\citeauthoryear{??}{mit}{}]%
        {mitmproxy}

\newblock \bibinfo{title}{{\texttt{mitmproxy}}}.
\newblock \bibinfo{howpublished}{\url{https://mitmproxy.org/}}.
\newblock
\newblock
\shownote{Online 06/18/2017.}


\bibitem[\protect\citeauthoryear{??}{pen}{}]%
        {penthouse}

\newblock \bibinfo{title}{{\texttt{penthouse}}}.
\newblock
  \bibinfo{howpublished}{\url{https://github.com/pocketjoso/penthouse}}.
\newblock
\newblock
\shownote{Online 06/18/2017.}


\bibitem[\protect\citeauthoryear{??}{spe}{}]%
        {speedindex}

\newblock \bibinfo{title}{{\texttt{SpeedIndex}}}.
\newblock
  \bibinfo{howpublished}{\newline{\url{https://sites.google.com/a/webpagetest.org/docs/using-webpagetest/metrics/speed-index}}}.
\newblock
\newblock
\shownote{Online 06/18/2017.}


\bibitem[\protect\citeauthoryear{Ager, Chatzis, Feldmann, Sarrar, Uhlig, and
  Willinger}{Ager et~al\mbox{.}}{2012}]%
        {SigcommIXP}
\bibfield{author}{\bibinfo{person}{Bernhard Ager}, \bibinfo{person}{Nikolaos
  Chatzis}, \bibinfo{person}{Anja Feldmann}, \bibinfo{person}{Nadi Sarrar},
  \bibinfo{person}{Steve Uhlig}, {and} \bibinfo{person}{Walter Willinger}.}
  \bibinfo{year}{2012}\natexlab{}.
\newblock \showarticletitle{Anatomy of a Large European IXP}. In
  \bibinfo{booktitle}{\emph{Proceedings of the ACM SIGCOMM 2012 Conference on
  Applications, Technologies, Architectures, and Protocols for Computer
  Communication}} \emph{(\bibinfo{series}{SIGCOMM '12})}.
  \bibinfo{publisher}{ACM}, \bibinfo{address}{New York, NY, USA},
  \bibinfo{pages}{163--174}.
\newblock


\bibitem[\protect\citeauthoryear{Archibald}{Archibald}{}]%
        {jake2017tough}
\bibfield{author}{\bibinfo{person}{Jake Archibald}.}
\newblock \bibinfo{title}{{HTTP/2 push is tougher than I thought}}.
\newblock
  \bibinfo{howpublished}{\url{https://jakearchibald.com/2017/h2-push-tougher-than-i-thought/}}.
\newblock
\newblock
\shownote{Online 06/18/2017.}


\bibitem[\protect\citeauthoryear{Belshe, Peon, and Thomson}{Belshe
  et~al\mbox{.}}{2015}]%
        {rfc7540}
\bibfield{author}{\bibinfo{person}{M. Belshe}, \bibinfo{person}{R. Peon}, {and}
  \bibinfo{person}{M. Thomson}.} \bibinfo{year}{2015}\natexlab{}.
\newblock \bibinfo{booktitle}{\emph{{Hypertext Transfer Protocol Version 2
  (HTTP/2)}}}.
\newblock \bibinfo{type}{RFC} 7540. \bibinfo{institution}{RFC Editor}.
\newblock
\showISSN{2070-1721}
\urldef\tempurl%
\url{http://www.rfc-editor.org/rfc/rfc7540.txt}
\showURL{%
\tempurl}


\bibitem[\protect\citeauthoryear{{Benedikt Wolters and Torsten
  Zimmermann}}{{Benedikt Wolters and Torsten Zimmermann}}{2018}]%
        {testbed}
\bibfield{author}{\bibinfo{person}{{Benedikt Wolters and Torsten Zimmermann}}.}
  \bibinfo{year}{2018}\natexlab{}.
\newblock \bibinfo{title}{Testbed Source and Measurement Results}.
\newblock
  \bibinfo{howpublished}{\url{https://github.com/COMSYS/http2-conext-push}}.
\newblock


\bibitem[\protect\citeauthoryear{Bergan, Pelchat, and Buettner}{Bergan
  et~al\mbox{.}}{}]%
        {thumb}
\bibfield{author}{\bibinfo{person}{Tom Bergan}, \bibinfo{person}{Simon
  Pelchat}, {and} \bibinfo{person}{Michael Buettner}.}
\newblock \bibinfo{title}{{Rules of Thumb for HTTP/2 Push}}.
\newblock
  \bibinfo{howpublished}{\url{https://docs.google.com/document/d/1K0NykTXBbbbTlv60t5MyJvXjqKGsCVNYHyLEXIxYMv0}}.
\newblock
\newblock
\shownote{Online 06/18/2017.}


\bibitem[\protect\citeauthoryear{Bocchi, De~Cicco, Mellia, and Rossi}{Bocchi
  et~al\mbox{.}}{2017}]%
        {bocchi2017web}
\bibfield{author}{\bibinfo{person}{Enrico Bocchi}, \bibinfo{person}{Luca
  De~Cicco}, \bibinfo{person}{Marco Mellia}, {and} \bibinfo{person}{Dario
  Rossi}.} \bibinfo{year}{2017}\natexlab{}.
\newblock \bibinfo{booktitle}{\emph{{The Web, the Users, and the MOS: Influence
  of HTTP/2 on User Experience}}}.
\newblock \bibinfo{publisher}{Springer International Publishing},
  \bibinfo{address}{Cham}, \bibinfo{pages}{47--59}.
\newblock


\bibitem[\protect\citeauthoryear{Borgnat, Dewaele, Fukuda, Abry, and
  Cho}{Borgnat et~al\mbox{.}}{2009}]%
        {SevenYears}
\bibfield{author}{\bibinfo{person}{P. Borgnat}, \bibinfo{person}{G. Dewaele},
  \bibinfo{person}{K. Fukuda}, \bibinfo{person}{P. Abry}, {and}
  \bibinfo{person}{K. Cho}.} \bibinfo{year}{2009}\natexlab{}.
\newblock \showarticletitle{Seven Years and One Day: Sketching the Evolution of
  Internet Traffic}. In \bibinfo{booktitle}{\emph{IEEE INFOCOM 2009}}.
  \bibinfo{pages}{711--719}.
\newblock


\bibitem[\protect\citeauthoryear{Butkiewicz, Madhyastha, and Sekar}{Butkiewicz
  et~al\mbox{.}}{2011}]%
        {butkiewicz2011complexity}
\bibfield{author}{\bibinfo{person}{Michael Butkiewicz},
  \bibinfo{person}{Harsha~V. Madhyastha}, {and} \bibinfo{person}{Vyas Sekar}.}
  \bibinfo{year}{2011}\natexlab{}.
\newblock \showarticletitle{{Understanding Website Complexity: Measurements,
  Metrics, and Implications}}. In \bibinfo{booktitle}{\emph{Proceedings of the
  2011 ACM SIGCOMM Conference on Internet Measurement Conference}}
  \emph{(\bibinfo{series}{IMC '11})}. \bibinfo{publisher}{ACM},
  \bibinfo{address}{New York, NY, USA}, \bibinfo{pages}{313--328}.
\newblock
\showISBNx{978-1-4503-1013-0}


\bibitem[\protect\citeauthoryear{Butkiewicz, Wang, Wu, Madhyastha, and
  Sekar}{Butkiewicz et~al\mbox{.}}{2015}]%
        {butkiewicz2015klotski}
\bibfield{author}{\bibinfo{person}{Michael Butkiewicz},
  \bibinfo{person}{Daimeng Wang}, \bibinfo{person}{Zhe Wu},
  \bibinfo{person}{Harsha~V. Madhyastha}, {and} \bibinfo{person}{Vyas Sekar}.}
  \bibinfo{year}{2015}\natexlab{}.
\newblock \showarticletitle{{Klotski: Reprioritizing Web Content to Improve
  User Experience on Mobile Devices}}. In \bibinfo{booktitle}{\emph{12th
  {USENIX} Symposium on Networked Systems Design and Implementation ({NSDI}
  15)}}. \bibinfo{publisher}{{USENIX} Association}, \bibinfo{address}{Oakland,
  CA}, \bibinfo{pages}{439--453}.
\newblock


\bibitem[\protect\citeauthoryear{de~Saxc{\'e}, Oprescu, and Chen}{de~Saxc{\'e}
  et~al\mbox{.}}{2015}]%
        {saxce2015h2}
\bibfield{author}{\bibinfo{person}{H. de Saxc{\'e}}, \bibinfo{person}{I.
  Oprescu}, {and} \bibinfo{person}{Y. Chen}.} \bibinfo{year}{2015}\natexlab{}.
\newblock \showarticletitle{{Is HTTP/2 really faster than HTTP/1.1?}}. In
  \bibinfo{booktitle}{\emph{2015 IEEE Conference on Computer Communications
  Workshops (INFOCOM WKSHPS)}}. \bibinfo{pages}{293--299}.
\newblock


\bibitem[\protect\citeauthoryear{Goel, Steiner, Wittie, Flack, and Ludin}{Goel
  et~al\mbox{.}}{2017}]%
        {goel2017measuring}
\bibfield{author}{\bibinfo{person}{Utkarsh Goel}, \bibinfo{person}{Moritz
  Steiner}, \bibinfo{person}{Mike~P. Wittie}, \bibinfo{person}{Martin Flack},
  {and} \bibinfo{person}{Stephen Ludin}.} \bibinfo{year}{2017}\natexlab{}.
\newblock \showarticletitle{{Measuring What is Not Ours: A Tale of 3rd Party
  Performance}}. In \bibinfo{booktitle}{\emph{Passive and Active Measurement}},
  \bibfield{editor}{\bibinfo{person}{Mohamed~Ali Kaafar},
  \bibinfo{person}{Steve Uhlig}, {and} \bibinfo{person}{Johanna Amann}} (Eds.).
  \bibinfo{publisher}{Springer International Publishing},
  \bibinfo{address}{Cham}, \bibinfo{pages}{142--155}.
\newblock


\bibitem[\protect\citeauthoryear{Grigorik}{Grigorik}{2013}]%
        {grigorik2013high}
\bibfield{author}{\bibinfo{person}{Ilya Grigorik}.}
  \bibinfo{year}{2013}\natexlab{}.
\newblock \bibinfo{booktitle}{\emph{{High Performance Browser Networking}}}.
\newblock \bibinfo{publisher}{{O'Reilly}}.
\newblock


\bibitem[\protect\citeauthoryear{Grigorik and Weiss}{Grigorik and Weiss}{}]%
        {grigorik2017preload}
\bibfield{author}{\bibinfo{person}{Ilya Grigorik} {and} \bibinfo{person}{Y.
  Weiss}.}
\newblock \bibinfo{title}{Preload}.
\newblock \bibinfo{howpublished}{\url{https://www.w3.org/TR/preload/}}.
\newblock
\newblock
\shownote{\newline{Online 06/18/2017}.}


\bibitem[\protect\citeauthoryear{Guercio}{Guercio}{}]%
        {cloudflare2018ab}
\bibfield{author}{\bibinfo{person}{Remy Guercio}.}
\newblock \bibinfo{title}{{Test New Features and Iterate Quickly with
  Cloudflare Workers}}.
\newblock
  \bibinfo{howpublished}{\url{https://blog.cloudflare.com/iterate-quickly-with-cloudflare-workers/}}.
\newblock
\newblock
\shownote{Online 10/03/2018.}


\bibitem[\protect\citeauthoryear{Han, Hao, and Qian}{Han et~al\mbox{.}}{2015}]%
        {han2015meta}
\bibfield{author}{\bibinfo{person}{Bo Han}, \bibinfo{person}{Shuai Hao}, {and}
  \bibinfo{person}{Feng Qian}.} \bibinfo{year}{2015}\natexlab{}.
\newblock \showarticletitle{{MetaPush: Cellular-Friendly Server Push For
  HTTP/2}}. In \bibinfo{booktitle}{\emph{Proceedings of the 5th Workshop on All
  Things Cellular: Operations, Applications and Challenges}}
  \emph{(\bibinfo{series}{AllThingsCellular '15})}. \bibinfo{publisher}{ACM},
  \bibinfo{address}{New York, NY, USA}, \bibinfo{pages}{57--62}.
\newblock
\showISBNx{978-1-4503-3538-6}


\bibitem[\protect\citeauthoryear{Jackel}{Jackel}{}]%
        {fastly2016ab}
\bibfield{author}{\bibinfo{person}{Chris Jackel}.}
\newblock \bibinfo{title}{{A/B testing at the edge}}.
\newblock
  \bibinfo{howpublished}{\url{https://www.fastly.com/blog/ab-testing-edge}}.
\newblock
\newblock
\shownote{Online 10/03/2018.}


\bibitem[\protect\citeauthoryear{Kelton, Ryoo, Balasubramanian, and Das}{Kelton
  et~al\mbox{.}}{2017}]%
        {kelton2017webgaze}
\bibfield{author}{\bibinfo{person}{Conor Kelton}, \bibinfo{person}{Jihoon
  Ryoo}, \bibinfo{person}{Aruna Balasubramanian}, {and}
  \bibinfo{person}{Samir~R. Das}.} \bibinfo{year}{2017}\natexlab{}.
\newblock \showarticletitle{{Improving User Perceived Page Load Times Using
  Gaze}}. In \bibinfo{booktitle}{\emph{14th {USENIX} Symposium on Networked
  Systems Design and Implementation ({NSDI} 17)}}. \bibinfo{publisher}{{USENIX}
  Association}, \bibinfo{address}{Boston, MA}, \bibinfo{pages}{545--559}.
\newblock


\bibitem[\protect\citeauthoryear{Lassey}{Lassey}{}]%
        {chrome2018ietf}
\bibfield{author}{\bibinfo{person}{Brad Lassey}.}
\newblock \bibinfo{title}{{Chrome's view on Push}}.
\newblock
  \bibinfo{howpublished}{\url{https://github.com/httpwg/wg-materials/blob/gh-pages/ietf102/chrome_push.pdf}}.
\newblock
\newblock
\shownote{Online 10/02/2018.}


\bibitem[\protect\citeauthoryear{Maier, Feldmann, Paxson, and Allman}{Maier
  et~al\mbox{.}}{2009}]%
        {GregorIMC}
\bibfield{author}{\bibinfo{person}{Gregor Maier}, \bibinfo{person}{Anja
  Feldmann}, \bibinfo{person}{Vern Paxson}, {and} \bibinfo{person}{Mark
  Allman}.} \bibinfo{year}{2009}\natexlab{}.
\newblock \showarticletitle{On Dominant Characteristics of Residential
  Broadband Internet Traffic}. In \bibinfo{booktitle}{\emph{Proceedings of the
  9th ACM SIGCOMM Conference on Internet Measurement}}
  \emph{(\bibinfo{series}{IMC '09})}. \bibinfo{publisher}{ACM},
  \bibinfo{address}{New York, NY, USA}, \bibinfo{pages}{90--102}.
\newblock
\showISBNx{978-1-60558-771-4}


\bibitem[\protect\citeauthoryear{Meenan}{Meenan}{2013}]%
        {meenan2013fast}
\bibfield{author}{\bibinfo{person}{Patrick Meenan}.}
  \bibinfo{year}{2013}\natexlab{}.
\newblock \showarticletitle{How Fast is Your Website?}
\newblock \bibinfo{journal}{\emph{Commun. ACM}} \bibinfo{volume}{56},
  \bibinfo{number}{4} (\bibinfo{date}{April} \bibinfo{year}{2013}),
  \bibinfo{pages}{49--55}.
\newblock
\showISSN{0001-0782}


\bibitem[\protect\citeauthoryear{Nanner}{Nanner}{}]%
        {akamai2018ietf}
\bibfield{author}{\bibinfo{person}{Aman Nanner}.}
\newblock \bibinfo{title}{{H2 Server Push Performance}}.
\newblock
  \bibinfo{howpublished}{\url{https://github.com/httpwg/wg-materials/blob/gh-pages/ietf102/akamai-server-push.pdf}}.
\newblock
\newblock
\shownote{Online 10/02/2018.}


\bibitem[\protect\citeauthoryear{Netravali, Sivaraman, Das, Goyal, Winstein,
  Mickens, and Balakrishnan}{Netravali et~al\mbox{.}}{2015}]%
        {netravali2015mahimahi}
\bibfield{author}{\bibinfo{person}{Ravi Netravali}, \bibinfo{person}{Anirudh
  Sivaraman}, \bibinfo{person}{Somak Das}, \bibinfo{person}{Ameesh Goyal},
  \bibinfo{person}{Keith Winstein}, \bibinfo{person}{James Mickens}, {and}
  \bibinfo{person}{Hari Balakrishnan}.} \bibinfo{year}{2015}\natexlab{}.
\newblock \showarticletitle{{Mahimahi: Accurate Record-and-Replay for {HTTP}}}.
  In \bibinfo{booktitle}{\emph{2015 {USENIX} Annual Technical Conference
  ({USENIX} {ATC} 15)}}. \bibinfo{publisher}{{USENIX} Association},
  \bibinfo{address}{Santa Clara, CA}, \bibinfo{pages}{417--429}.
\newblock


\bibitem[\protect\citeauthoryear{Nottingham}{Nottingham}{}]%
        {pushissue}
\bibfield{author}{\bibinfo{person}{Mark Nottingham}.}
\newblock \bibinfo{title}{{httpwg: Issue \#579}}.
\newblock
  \bibinfo{howpublished}{\url{https://github.com/httpwg/http-extensions/issues/579}}.
\newblock
\newblock
\shownote{Online 06/18/2017.}


\bibitem[\protect\citeauthoryear{Oku and Nottingham}{Oku and
  Nottingham}{2017}]%
        {oku2017cache}
\bibfield{author}{\bibinfo{person}{Kazuho Oku} {and} \bibinfo{person}{Mark
  Nottingham}.} \bibinfo{year}{2017}\natexlab{}.
\newblock \bibinfo{booktitle}{\emph{Cache Digests for HTTP/2}}.
\newblock \bibinfo{type}{Internet-Draft} draft-ietf-httpbis-cache-digest-02.
  \bibinfo{institution}{IETF Secretariat}.
\newblock
\urldef\tempurl%
\url{http://www.ietf.org/internet-drafts/draft-ietf-httpbis-cache-digest-02.txt}
\showURL{%
\tempurl}


\bibitem[\protect\citeauthoryear{Peon and Ruellan}{Peon and Ruellan}{2015}]%
        {rfc7541}
\bibfield{author}{\bibinfo{person}{R. Peon} {and} \bibinfo{person}{H.
  Ruellan}.} \bibinfo{year}{2015}\natexlab{}.
\newblock \bibinfo{booktitle}{\emph{HPACK: Header Compression for HTTP/2}}.
\newblock \bibinfo{type}{RFC} 7541. \bibinfo{institution}{RFC Editor}.
\newblock
\showISSN{2070-1721}
\urldef\tempurl%
\url{http://www.rfc-editor.org/rfc/rfc7541.txt}
\showURL{%
\tempurl}


\bibitem[\protect\citeauthoryear{Rosen, Han, Hao, Mao, and Qian}{Rosen
  et~al\mbox{.}}{2017}]%
        {rosen2017push}
\bibfield{author}{\bibinfo{person}{Sanae Rosen}, \bibinfo{person}{Bo Han},
  \bibinfo{person}{Shuai Hao}, \bibinfo{person}{Z.~Morley Mao}, {and}
  \bibinfo{person}{Feng Qian}.} \bibinfo{year}{2017}\natexlab{}.
\newblock \showarticletitle{{Push or Request: An Investigation of HTTP/2 Server
  Push for Improving Mobile Performance}}. In
  \bibinfo{booktitle}{\emph{Proceedings of the 26th International Conference on
  World Wide Web}} \emph{(\bibinfo{series}{WWW '17})}.
  \bibinfo{publisher}{International World Wide Web Conferences Steering
  Committee}, \bibinfo{address}{Republic and Canton of Geneva, Switzerland},
  \bibinfo{pages}{459--468}.
\newblock


\bibitem[\protect\citeauthoryear{Ruamviboonsuk, Netravali, Uluyol, and
  Madhyastha}{Ruamviboonsuk et~al\mbox{.}}{2017}]%
        {ruamviboonsuk2017vroom}
\bibfield{author}{\bibinfo{person}{Vaspol Ruamviboonsuk}, \bibinfo{person}{Ravi
  Netravali}, \bibinfo{person}{Muhammed Uluyol}, {and}
  \bibinfo{person}{Harsha~V. Madhyastha}.} \bibinfo{year}{2017}\natexlab{}.
\newblock \showarticletitle{{Vroom: Accelerating the Mobile Web with
  Server-Aided Dependency Resolution}}. In
  \bibinfo{booktitle}{\emph{Proceedings of the Conference of the ACM Special
  Interest Group on Data Communication}} \emph{(\bibinfo{series}{SIGCOMM
  '17})}. \bibinfo{publisher}{ACM}, \bibinfo{address}{New York, NY, USA},
  \bibinfo{pages}{390--403}.
\newblock


\bibitem[\protect\citeauthoryear{R{\"u}th, Poese, Dietzel, and
  Hohlfeld}{R{\"u}th et~al\mbox{.}}{2018}]%
        {rueth2018quic}
\bibfield{author}{\bibinfo{person}{Jan R{\"u}th}, \bibinfo{person}{Ingmar
  Poese}, \bibinfo{person}{Christoph Dietzel}, {and} \bibinfo{person}{Oliver
  Hohlfeld}.} \bibinfo{year}{2018}\natexlab{}.
\newblock \showarticletitle{A First Look at QUIC in the Wild}. In
  \bibinfo{booktitle}{\emph{Passive and Active Measurement}},
  \bibfield{editor}{\bibinfo{person}{Robert Beverly}, \bibinfo{person}{Georgios
  Smaragdakis}, {and} \bibinfo{person}{Anja Feldmann}} (Eds.).
  \bibinfo{publisher}{Springer International Publishing},
  \bibinfo{address}{Cham}, \bibinfo{pages}{255--268}.
\newblock


\bibitem[\protect\citeauthoryear{Scheitle, Hohlfeld, Gamba, Jelten, Zimmermann,
  Strowes, and Vallina-Rodriguez}{Scheitle et~al\mbox{.}}{2018}]%
        {scheitle2018toplist}
\bibfield{author}{\bibinfo{person}{Quirin Scheitle}, \bibinfo{person}{Oliver
  Hohlfeld}, \bibinfo{person}{Julien Gamba}, \bibinfo{person}{Jonas Jelten},
  \bibinfo{person}{Torsten Zimmermann}, \bibinfo{person}{Stephen~D. Strowes},
  {and} \bibinfo{person}{Narseo Vallina-Rodriguez}.}
  \bibinfo{year}{2018}\natexlab{}.
\newblock \showarticletitle{A Long Way to the Top: Significance, Structure, and
  Stability of Internet Top Lists}. In \bibinfo{booktitle}{\emph{Proceedings of
  the 2018 Internet Measurement Conference}} \emph{(\bibinfo{series}{IMC
  '18})}. \bibinfo{publisher}{ACM}, \bibinfo{address}{New York, NY, USA}.
\newblock


\bibitem[\protect\citeauthoryear{Varvello, Schomp, Naylor, Blackburn, Finamore,
  and Papagiannaki}{Varvello et~al\mbox{.}}{2016}]%
        {varvello2016web}
\bibfield{author}{\bibinfo{person}{Matteo Varvello}, \bibinfo{person}{Kyle
  Schomp}, \bibinfo{person}{David Naylor}, \bibinfo{person}{Jeremy Blackburn},
  \bibinfo{person}{Alessandro Finamore}, {and} \bibinfo{person}{Konstantina
  Papagiannaki}.} \bibinfo{year}{2016}\natexlab{}.
\newblock \showarticletitle{{Is the Web HTTP/2 Yet?}}. In
  \bibinfo{booktitle}{\emph{International Conference on Passive and Active
  Network Measurement}}. Springer, \bibinfo{pages}{218--232}.
\newblock


\bibitem[\protect\citeauthoryear{Wang, Balasubramanian, Krishnamurthy, and
  Wetherall}{Wang et~al\mbox{.}}{2013}]%
        {wang2013wprof}
\bibfield{author}{\bibinfo{person}{Xiao~Sophia Wang}, \bibinfo{person}{Aruna
  Balasubramanian}, \bibinfo{person}{Arvind Krishnamurthy}, {and}
  \bibinfo{person}{David Wetherall}.} \bibinfo{year}{2013}\natexlab{}.
\newblock \showarticletitle{{Demystifying Page Load Performance with WProf}}.
  In \bibinfo{booktitle}{\emph{Presented as part of the 10th {USENIX} Symposium
  on Networked Systems Design and Implementation ({NSDI} 13)}}.
  \bibinfo{publisher}{{USENIX}}, \bibinfo{address}{Lombard, IL},
  \bibinfo{pages}{473--485}.
\newblock


\bibitem[\protect\citeauthoryear{Wang, Balasubramanian, Krishnamurthy, and
  Wetherall}{Wang et~al\mbox{.}}{2014}]%
        {wang2014spdy}
\bibfield{author}{\bibinfo{person}{Xiao~Sophia Wang}, \bibinfo{person}{Aruna
  Balasubramanian}, \bibinfo{person}{Arvind Krishnamurthy}, {and}
  \bibinfo{person}{David Wetherall}.} \bibinfo{year}{2014}\natexlab{}.
\newblock \showarticletitle{{How Speedy is SPDY?}}. In
  \bibinfo{booktitle}{\emph{Proceedings of the 11th USENIX Conference on
  Networked Systems Design and Implementation}}
  \emph{(\bibinfo{series}{NSDI'14})}. \bibinfo{publisher}{USENIX Association},
  \bibinfo{address}{Berkeley, CA, USA}, \bibinfo{pages}{387--399}.
\newblock


\bibitem[\protect\citeauthoryear{Zarifis, Holland, Jain, Katz-Bassett, and
  Govindan}{Zarifis et~al\mbox{.}}{2017}]%
        {zafiris2017making}
\bibfield{author}{\bibinfo{person}{Kyriakos Zarifis}, \bibinfo{person}{Mark
  Holland}, \bibinfo{person}{Manish Jain}, \bibinfo{person}{Ethan
  Katz-Bassett}, {and} \bibinfo{person}{Ramesh Govindan}.}
  \bibinfo{year}{2017}\natexlab{}.
\newblock \bibinfo{booktitle}{\emph{{Making Effective Use of HTTP/2 Server Push
  in Content Delivery Networks}}}.
\newblock \bibinfo{type}{{T}echnical {R}eport}.
  \bibinfo{institution}{University of Southern California}.
\newblock


\bibitem[\protect\citeauthoryear{Zimmermann, R{\"u}th, Wolters, and
  Hohlfeld}{Zimmermann et~al\mbox{.}}{2017a}]%
        {zimmermann2017push}
\bibfield{author}{\bibinfo{person}{Torsten Zimmermann}, \bibinfo{person}{Jan
  R{\"u}th}, \bibinfo{person}{Benedikt Wolters}, {and} \bibinfo{person}{Oliver
  Hohlfeld}.} \bibinfo{year}{2017}\natexlab{a}.
\newblock \showarticletitle{{How HTTP/2 Pushes the Web: An Empirical Study of
  HTTP/2 Server Push}}. In \bibinfo{booktitle}{\emph{2017 IFIP Networking
  Conference (IFIP Networking) and Workshops}}.
\newblock


\bibitem[\protect\citeauthoryear{Zimmermann, Wolters, and Hohlfeld}{Zimmermann
  et~al\mbox{.}}{2017b}]%
        {zimmermann2017pushqoe}
\bibfield{author}{\bibinfo{person}{Torsten Zimmermann},
  \bibinfo{person}{Benedikt Wolters}, {and} \bibinfo{person}{Oliver Hohlfeld}.}
  \bibinfo{year}{2017}\natexlab{b}.
\newblock \showarticletitle{{A QoE Perspective on HTTP/2 Server Push}}. In
  \bibinfo{booktitle}{\emph{Proceedings of the Workshop on QoE-based Analysis
  and Management of Data Communication Networks}}
  \emph{(\bibinfo{series}{Internet QoE '17})}. \bibinfo{publisher}{ACM},
  \bibinfo{address}{New York, NY, USA}, \bibinfo{pages}{1--6}.
\newblock


\end{thebibliography}

\end{document}